\documentclass[12pt,preprint]{aastex}
\received{nnn}
\revised{nnn}
\accepted{nnn}
\shorttitle{NGC 7097: the AGN and its mirror.}
\shortauthors{Ricci et al.}

\begin{document}

\title{NGC 7097: the AGN and its mirror, revealed by PCA  Tomography}
\author{T.V. Ricci$^{\dagger}$, J.E. Steiner \& R.B. Menezes}
\affil{Instituto de Astronomia Geof\'isica e Ci\^encias Atmosf\'ericas - Universidade de S\~ao Paulo}
\affil{Rua do Mat\~ao, 1226, Cidade Universit\'aria, S\~ao Paulo - SP, Brazil CEP 05508-090}
\email{$^\dagger$tiago@astro.iag.usp.br}
\begin{abstract}

Three-dimensional (3D) spectroscopy techniques are becoming more and more popular, producing an increasing number of large data cubes. The challenge of extracting information from these cubes requires the development of new techniques for data processing and analysis. We apply the recently developed technique of Principal Component Analysis (PCA) Tomography to a data cube from the center of the elliptical galaxy NGC 7097 and show that this technique is effective in decomposing the data into physically interpretable information. We find that the first five principal components of our data are associated with distinct physical characteristics. In particular, we detect a LINER with a weak broad component in the Balmer lines. Two images of the LINER are present in our data, one seen through a disk of gas and dust, and the other after scattering by free electrons and/or dust particles in the ionization cone. Furthermore, we extract the spectrum of the LINER, decontaminated from stellar and extended nebular emission, using only the technique of PCA Tomography. We anticipate that the scattered image has polarized light, due to its scattered nature.

\end{abstract}

\keywords{Galaxies: active - galaxies: elliptical and lenticular, cD - galaxies: individual (NGC 7097) – galaxies: kinematics and dynamics - galaxies: nuclei - techniques: spectroscopic}

\section{INTRODUCTION} \label{introduction}

Active Galactic Nuclei (AGN) form a heterogeneous group of objects including quasars, Seyfert galaxies, LINERS and others. The dominant view today is  that these objects are associated with the capture of gas by a supermassive black hole located at the center of those galaxies. In the optical, their spectra are characterized by broad and intense emission lines. Quasars are among the more luminous types of AGN and are typically found at higher redshifts. At lower luminosities one finds Seyfert galaxies, whose spectra are characterized by intense lines from high ionization species  and LINERs (Low Ionization Nuclear Emitting Regions - \citealt{1980A&A....87..152H}), with enhanced lines from low ionization species. In both cases, the host galaxy is usually detected. Seyfert galaxies are typically spiral galaxies, while LINERs and quasars are more frequently seen in elliptical galaxies \citep{2008ARA&A..46..475H}.

Seyfert galaxies (and LINERs as well) can be divided into two types: those that have permitted lines wider than the forbidden lines (type 1) and those that have both types of lines with similar widths (type 2). The unified model \citep{1978PNAS...75..540O,1993ARA&A..31..473A} suggests that the broad line emission region (BLR) is surrounded by a torus of gas and dust; so depending on the orientation of this torus relative to our line of sight, one can or not observe the BLR, thereby explaining the existence of the two types of Seyfert galaxies. This model is convincingly verified by the spectropolarimetric observations of NGC 1068 \citep{1985ApJ...297..621A}.

In this article, we analyze a data cube for the central part of the elliptical galaxy NGC 7097, obtained with the GMOS-IFU spectrograph on the Gemini South Telescope, with the goal of detecting and characterizing low-luminosity AGNs in early type galaxies. Data cubes are large sets of measurements with two spatial (spaxels) and one spectral (pixels) dimension, composed of several tens of millions of measurements. Extracting information from such cubes is a complex task and the traditional methods of spectral analysis are inadequate, so new methods are much needed and welcome. The number of spectrographs capable of producing such large data cubes is increasing rapidly. For instance, the JWST (James Webb Space Telescope) will have 5 IFUs and all projected ELTs (Extremely Large Telescopes) plan to provide at least one IFU in their first generation instruments. In this paper, we analyze a data cube of NGC 7097 with a recently developed methodology (PCA Tomography; \citealt{2009MNRAS.395...64S}) and show its power to extract useful information.

NGC 7097 is an E5 elliptical galaxy, according to RC 3 \citep{1991rc3..book.....D} at a distance of 32.4 Mpc \citep{2001ApJ...546..681T}. This galaxy shows strong emission of [O II] $\lambda$3727\AA\ \citep{1984PASP...96..287C} and [N II] $\lambda$6584\AA\ which has a width of FWHM = 423 km/ s \citep{1986AJ.....91.1062P}; the strong [N II] emission was confirmed by \citet{1997A&A...323..349P}. The presence of these strong low ionization emission lines, not produced in H II regions, indicates that the gas requires a non-thermal ionization source. \citet{1986ApJ...305..136C} detected a disk of gas, with a radius of 15'', aligned with the axis of the galaxy (the position angle of the galaxy is $PA=18^o$), counter-rotating with respect to the stellar component. As the extent of the galaxy is 1.9 arcmin on the sky, the gaseous disk has nearly $\frac{1}{4}$ of the size of the galaxy. However, \citet{1993A&A...280..409B}, using imaging in H$\alpha$+[N II] and the adjacent continuum, found that the extended emission is rotated by $-30^o$ from the stellar isophotes, that is, the emission disk has a $PA=-12^o$. These authors estimate the inclination of the gaseous disk as $57^o$. \citet{1996A&AS..120..257Z} also detected a disk structure with an inclination $i = 62^o$ with respect to the plane of the sky while \citet{1997A&A...323..349P} found $67^o$. 

\section{OBSERVATIONS, REDUCTION AND DATA TREATMENT} \label {treatment}

The observations of NGC 7097 were made on July 31, 2007 at the Gemini South Telescope \footnote{Based on observations obtained at the Gemini Observatory, which is operated by the
Association of Universities for Research in Astronomy, Inc., under a cooperative agreement
with the NSF on behalf of the Gemini partnership: the National Science Foundation (United
States), the Science and Technology Facilities Council (United Kingdom), the
National Research Council (Canada), CONICYT (Chile), the Australian Research Council
(Australia), Minist\'erio da Ci\^encia e Tecnologia (Brazil) 
and Ministerio de Ciencia, Tecnolog\'ia e Innovaci\'on Productiva  (Argentina)}. We used the ``Integral Field Unit'' - IFU of the Gemini Multi Object Spectrograph - GMOS \citep{2004PASP..116..425H,2002PASP..114..892A} in the single slit mode. This mode allows the object and the sky to be observed simultaneously by 500 and 250 micro-lenses, respectively, located in the focal plane of the telescope. The micro-lenses divide the image of the object into slices of 0.2 arcseconds and is coupled to an array of optical fibers arranged linearly on the nominal location of the slit spectrograph (pseudo-gap). The observation of the object and sky are separated by one arcminute. The final product is a data cube with two spatial dimensions spanning 3.5 x 5 arcsec and one spectral dimension. This allows the construction of images in a specific wavelength range, or  the extraction of the spectrum in a given region of space. We use the grating B600-G5323, with a central wavelength of observation in 5650\AA . The spectra cover a range of 4228 - 7120\AA\ and have a resolution of 1.8\AA , as measured from the skyline [O I] 5577\AA . 

The usual lamp and sky flat-fields as well as bias exposures were taken for the data cube corrections. Exposures of a Cu-Ar lamp were taken for wavelength calibration and a data cube of the star LTT 9239 was obtained for the flux calibration. The seeing was measured to be 1.0''. The resulting data were reduced using the standard Gemini IRAF\footnote{IRAF is distributed by the National Optical Astronomy Observatory, which is operated by the Association of Universities for Research in Astronomy (AURA) under cooperative agreement with the National Science Foundation.} package and a spatial sampling of 0.05 arcsec per pixel, while the CCD spaxel has 0.1'' x 0.1''. Bias and background subtractions were made, as well as the corrections for the fiber response, removal of cosmic rays and subtraction of the sky from both the observations of the galaxy and also from the standard star.

The spatial high frequency noise in the data cube was suppressed using a Butterworth filter \citep{2002dip..book.....G} applied to the Fourier transform of each image in the data cube. The parameters of the filter $H(u,v)$ were $n=6$ and $a=b=0.15 F_{NY}$, i.e. $H(a,b)=0.5$, where $F_{NY}$ is the Nyquist frequency of the spatial component of the cube. After removing the high frequency noise, the data cube of NGC 7097 still has a signature of low-frequency instrumental noise, which is present in almost all the observations made with the GMOS-IFU. To reduce such instrumental ``fingerprints'' we have developed a technique that uses PCA Tomography in the wavelet space of the data cube's spatial dimensions. We have no space here to describe this procedure, nor is it the scope of this letter. This technique is laborious and is described in detail in Steiner et al. (2011, in preparation). This technique was applied to successfully remove the low frequency instrumental noise to the data cube discussed in this Letter.

The GMOS-IFU data were obtained without an atmospheric dispersion corrector (ADC). Thus, the data cube suffers from the effect of atmospheric dispersion, which increases with the air mass of the observation. To correct for the atmospheric dispersion, we used the equations of \citet{1982PASP...94..715F} and \citet{1998Metro..35..133B}. Finally, the cube was deconvolved in the spatial dimension using the Richardson-Lucy method with 6 iterations and a Gaussian PSF with FWHM = 1.0''. After deconvolution, the estimated final PSF has a FWHM = 0.7'', as indicated by our experience from processing many other stellar profiles of other observing programs.

\section{DATA ANALYSIS AND RESULTS} \label{results}

The analysis of large data cubes may become complex and overwhelming, as it may involve tens of millions of pixels. More concerning is that, given this complexity, only some restricted subset of the data ends up being analyzed; the rest is at the risk of being largely ignored. To overcome the complexity of analyzing such large data cubes, we employ a method of data cube analysis based on Principal Component Analysis (PCA). This method extracts the significant information content associated with the data through an effective dimensional reduction, facilitating its interpretation. PCA compresses the data, originally expressed as a large set of correlated variables, into a small but optimal set of uncorrelated variables, ordered by their eigenvalues. An important aspect of PCA is that the eigenvectors are mutually orthogonal; if there are physically uncorrelated phenomena expressed in the data cube, they possibly will align with different eigenvectors in decreasing order of importance. Furthermore, PCA is a nonparametric statistic, therefore there are no parameters or coefficients to adjust that somehow depend on the users' experience and skills, or on physical and geometrical parameters of a proposed model. PCA, therefore, provides a unique and objective solution.  

PCA has been used many times in the astronomical literature, and a more extended presentation of this technique is given in \citet{1987ASSL..131.....M} and \citet{fukunaga}. Most of the applications of PCA in astronomy are related to finding eigenvectors for a population of objects. In the present case, we want to apply the technique to a single data cube in which the objects are spaxels (spatial pixels). The wavelength pixels are their properties. PCA produces eigenvectors (the uncorrelated variables), which we refer to, also, as eigenspectra, and tomograms, which are images of the data projected onto each eigenvector. In traditional tomographic techniques, one obtains images that represent `slices' in three-dimensional space (the human body, for example), or in velocity space (Doppler Tomography). In PCA Tomography, one obtains images that represent `slices' of the data (tomograms) in the eigenvector space. Each tomogram is associated with an eigenspectrum. The simultaneous analysis of both brings a new perspective to the interpretation of them. For a full presentation of PCA Tomography, see \citet{2009MNRAS.395...64S}; a shorter version is presented in \citet{2010IAUS..267...85S}.

\begin{figure}[h!]
\begin{center} 
\begin{minipage}[b]{0.50 \linewidth}
\begin{center}
\includegraphics[height=5.00cm,width=3.40cm]{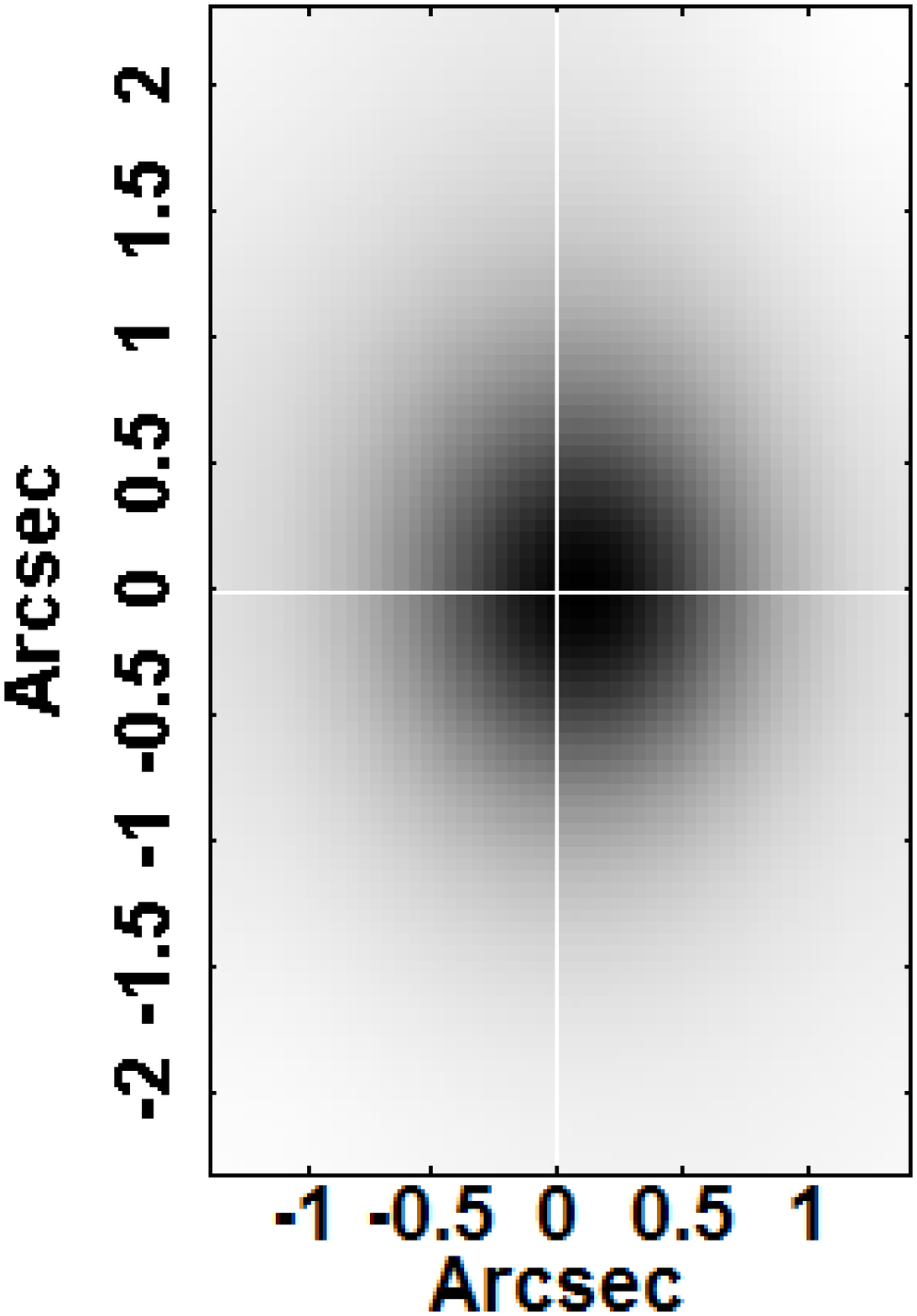}
\end{center}
\end{minipage}\hfill
\begin{minipage}[b]{0.50 \linewidth}
\includegraphics[height=5.00cm,width=5.40cm]{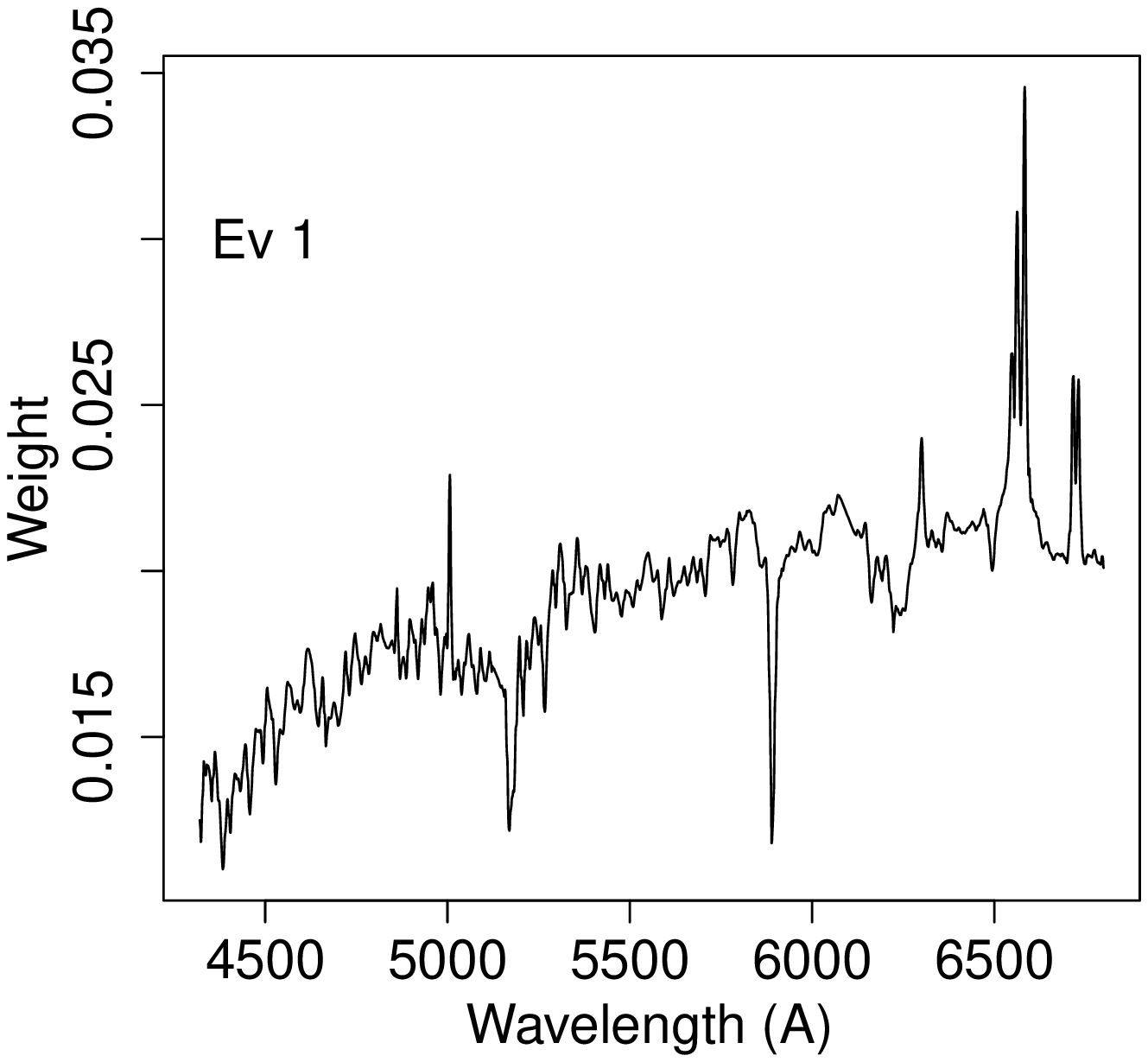}
\end{minipage}\hfill
\caption{Eigenvector (Ev 1) and tomogram of principal component 1 of the data cube of NGC 7097. Black means stronger and white, weaker correlation} 
\label{Eigenvec1_fig}
\end{center}
\end{figure}

\begin{figure}[h!]
	\begin{center}
	\includegraphics[height=8cm, width=8cm, angle=0]{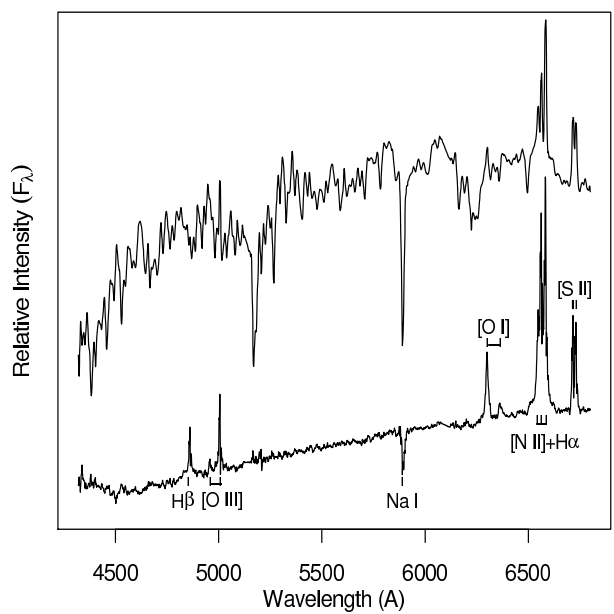}
	\caption{Spectrum of NGC 7097. The top curve shows the spectrum summed over all spaxels. This spectrum is dominated by the stellar emission in the field of view, but we can see the lines of H$\alpha$, [N II] and [S II]. The bottom curve shows the extracted AGN spectrum, using the technique of feature suppression. In this spectrum we can clearly see the features of LINER emission and broad, although weak, Balmer lines. The two spectra are shown in relative intensity units, but on the same scale.}
	\label{espec_total_av2eav4_N7097}
	\end{center}
\end{figure}

\begin{figure}[h!]
\begin{center} 
\begin{minipage}[b]{0.50 \linewidth}
\begin{center}
\includegraphics[height=5.00cm,width=3.40cm]{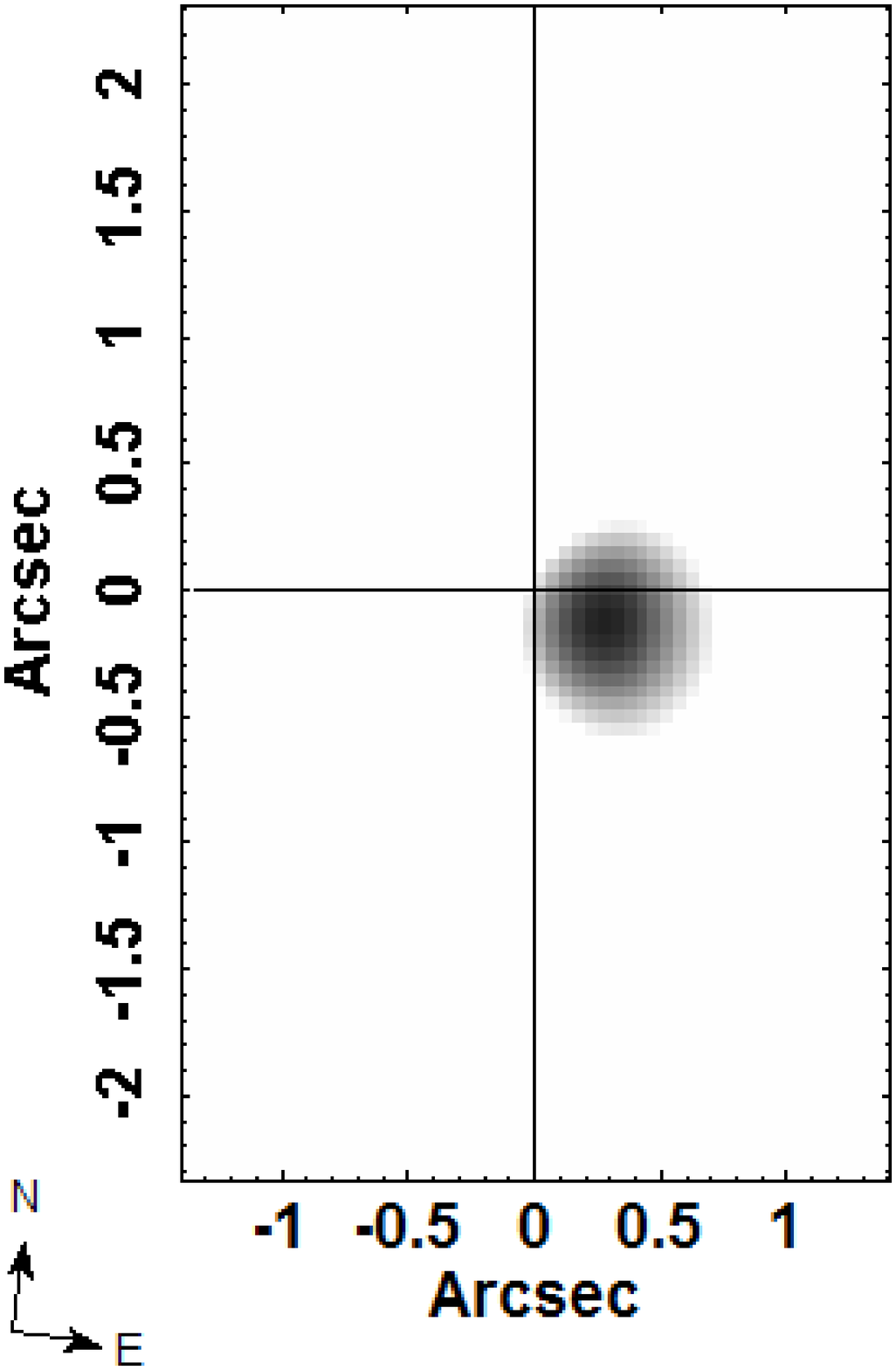}
\end{center}
\end{minipage}\hfill
\begin{minipage}[b]{0.50 \linewidth}
\includegraphics[height=5.00cm,width=5.40cm]{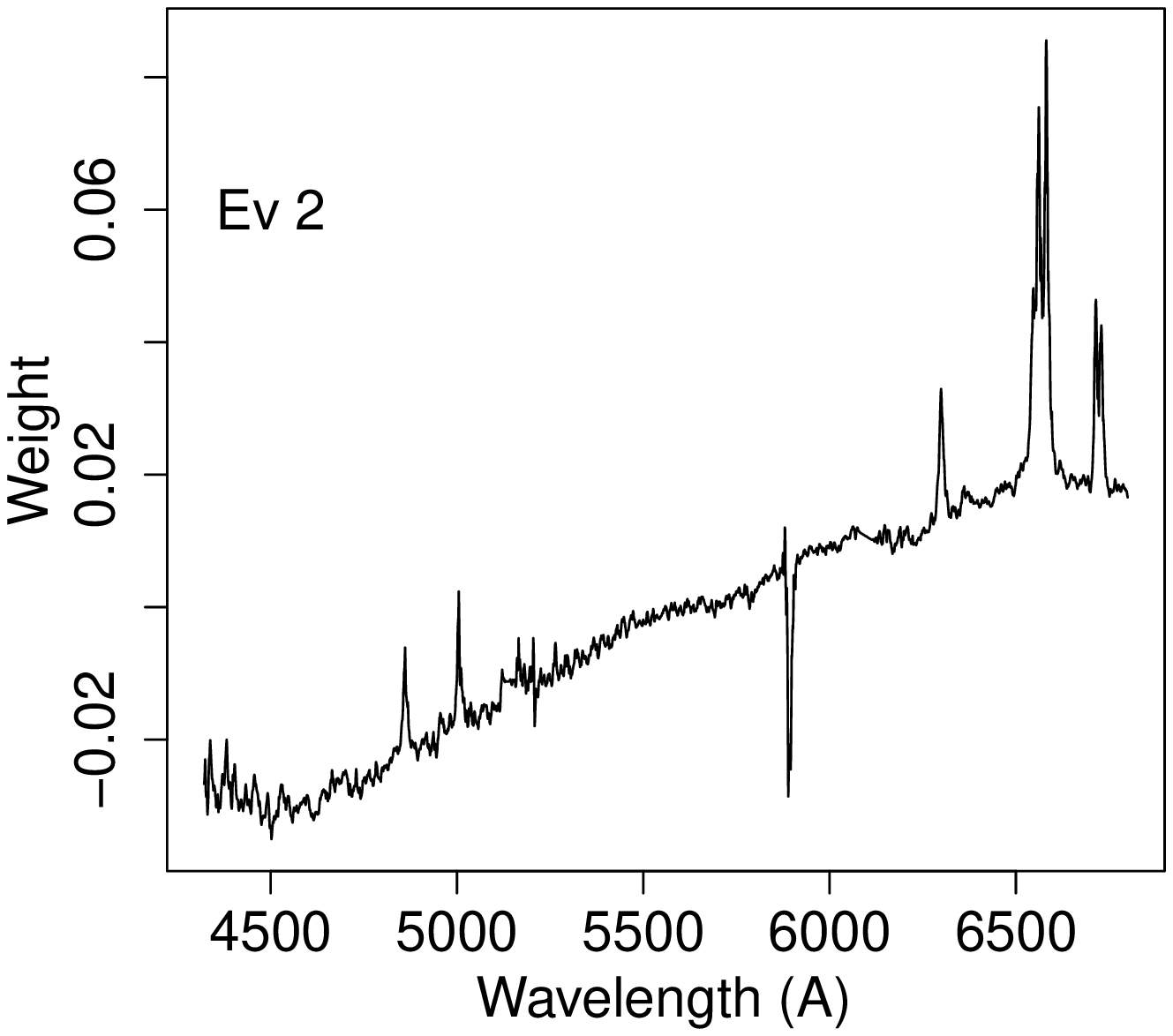}
\end{minipage}\hfill
\begin{minipage}[b]{0.50 \linewidth}
\begin{center}
\includegraphics[height=5.00cm,width=3.40cm]{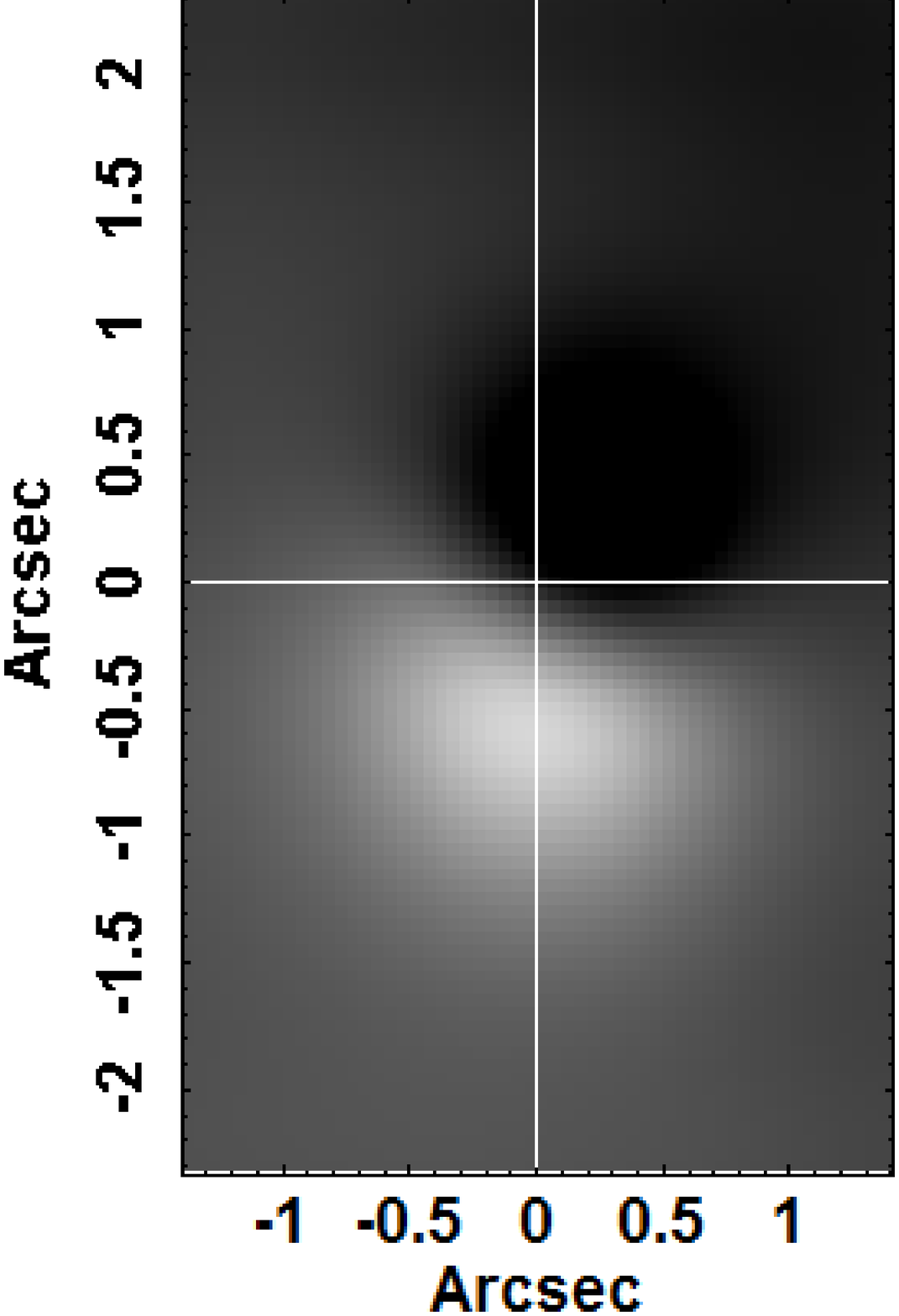}
\end{center}
\end{minipage}\hfill
\begin{minipage}[b]{0.50 \linewidth}
\includegraphics[height=5.00cm,width=5.40cm]{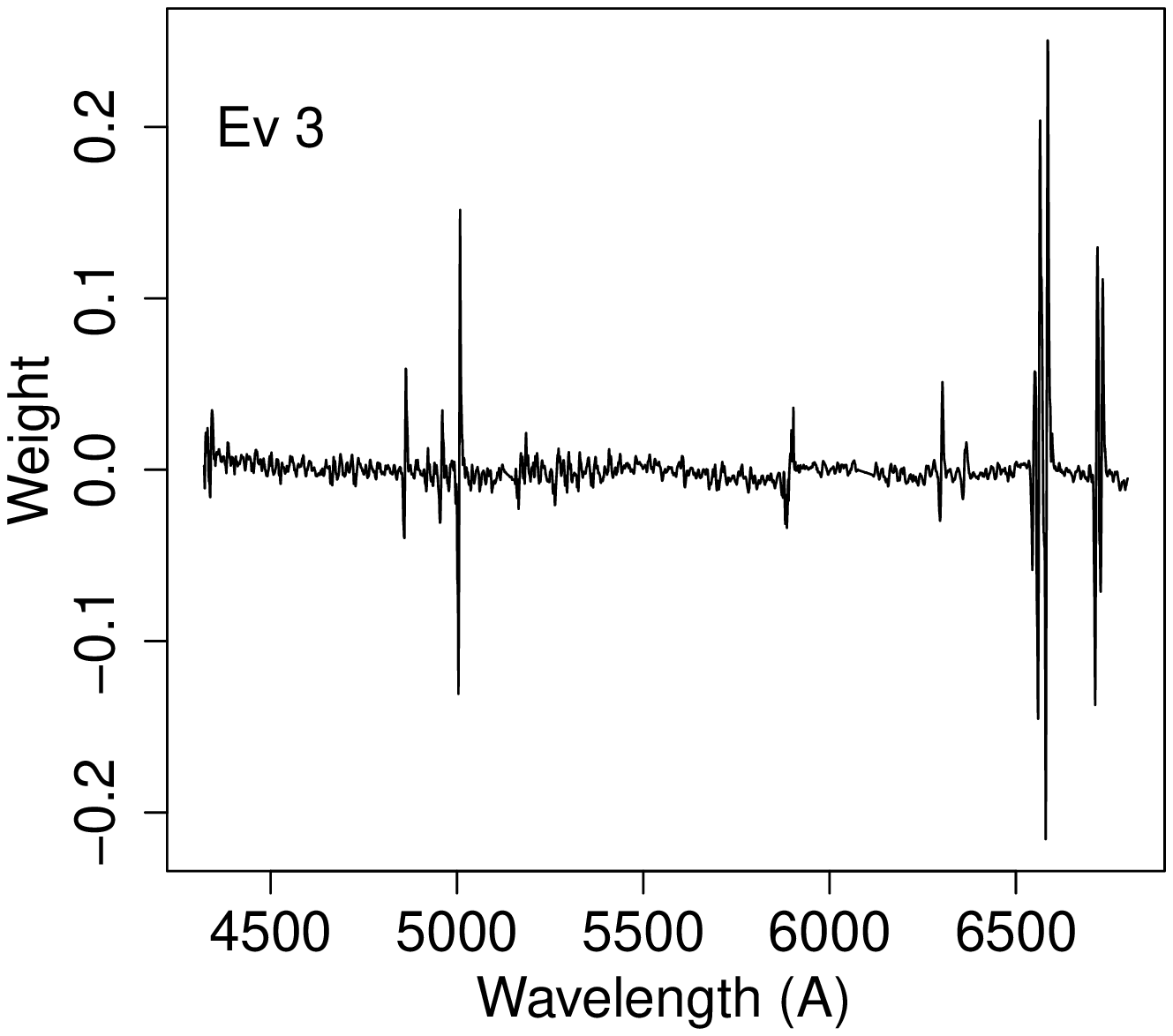}
\end{minipage}\hfill
\begin{minipage}[b]{0.50 \linewidth}
\begin{center}
\includegraphics[height=5.00cm,width=3.40cm]{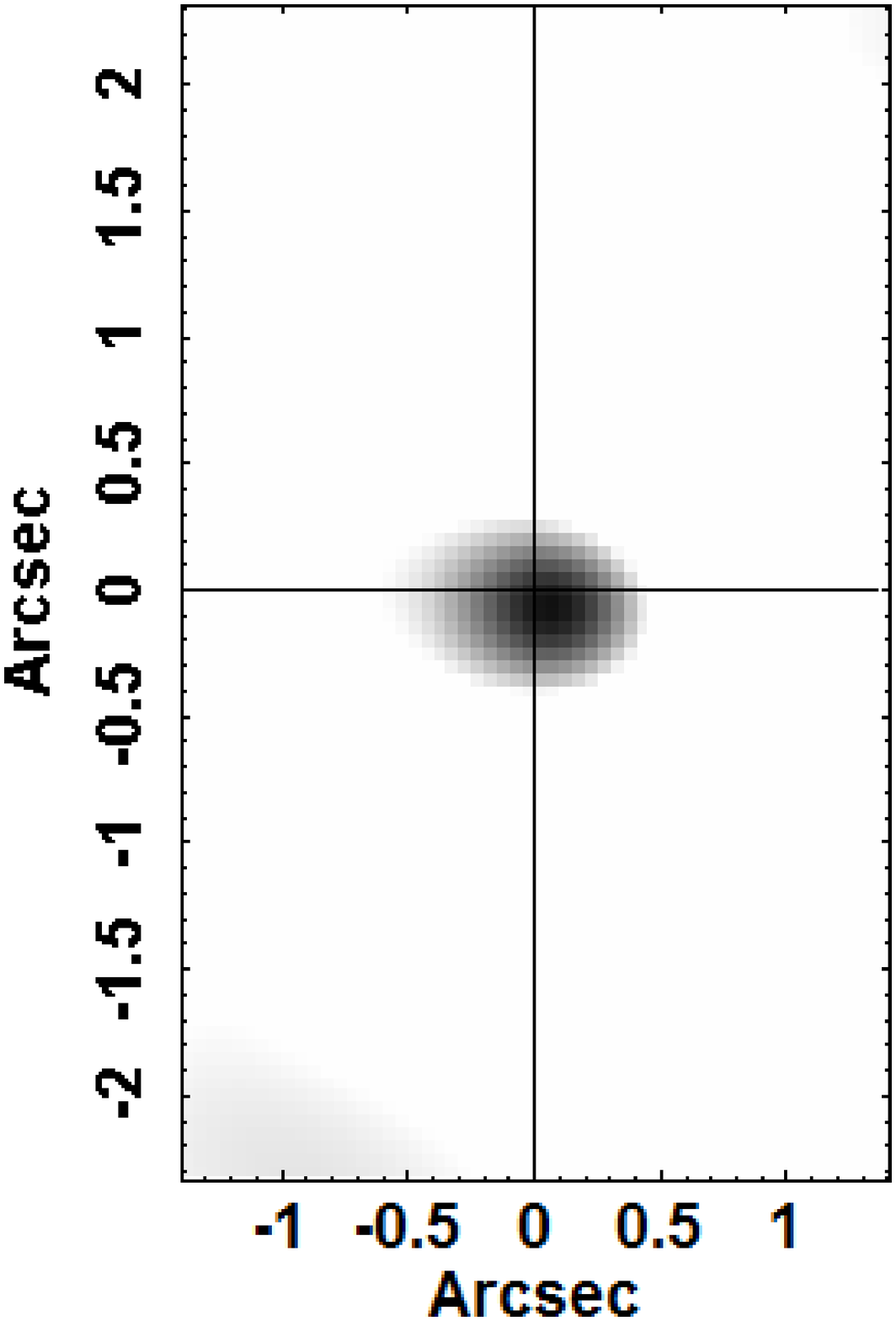}
\end{center}
\end{minipage}\hfill
\begin{minipage}[b]{0.50 \linewidth}
\includegraphics[height=5.00cm,width=5.40cm]{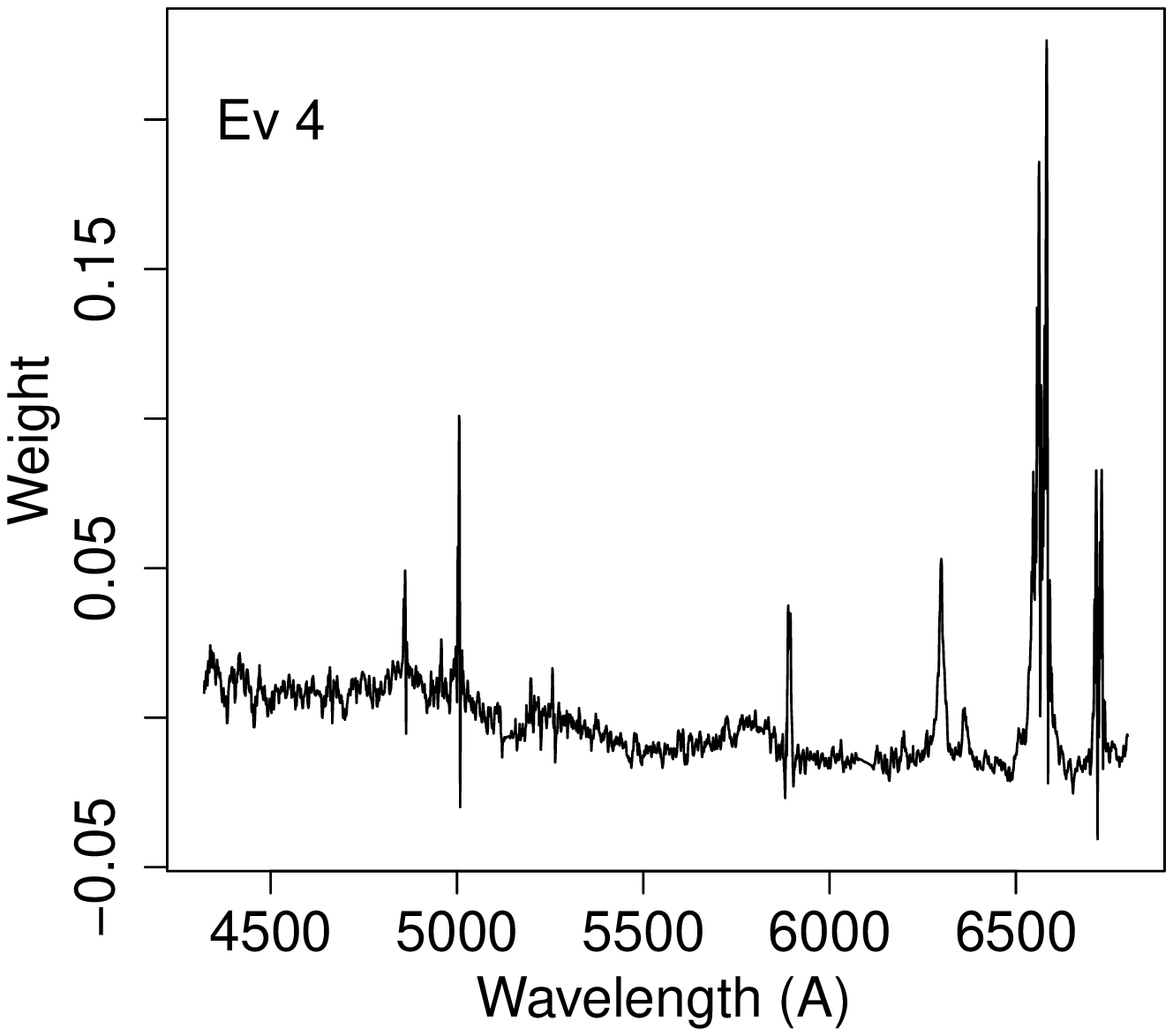}
\end{minipage}\hfill
\begin{minipage}[b]{0.50 \linewidth}
\begin{center}
\includegraphics[height=5.00cm,width=3.40cm]{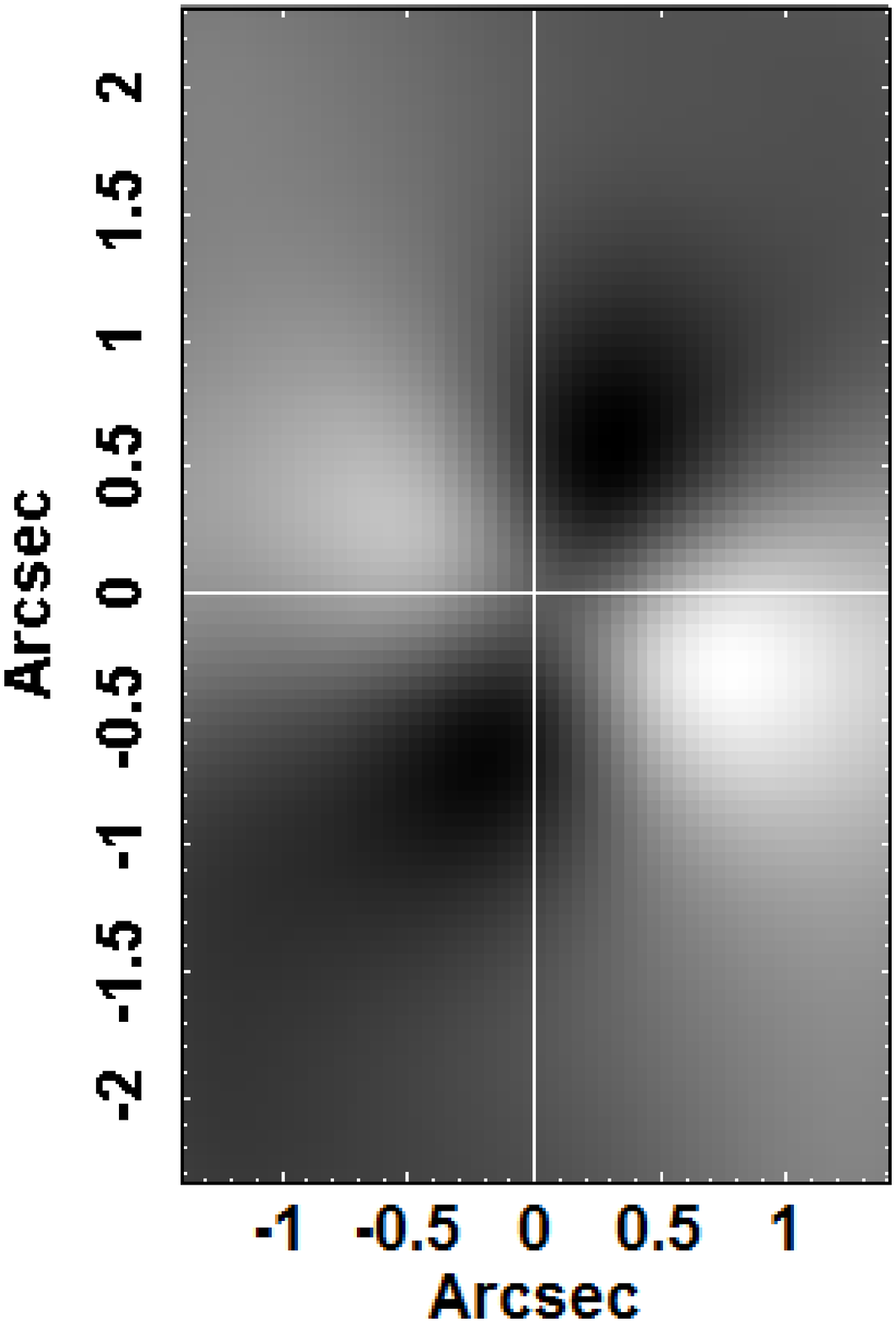}
\end{center}
\end{minipage}\hfill
\begin{minipage}[b]{0.50 \linewidth}
\includegraphics[height=5.00cm,width=5.40cm]{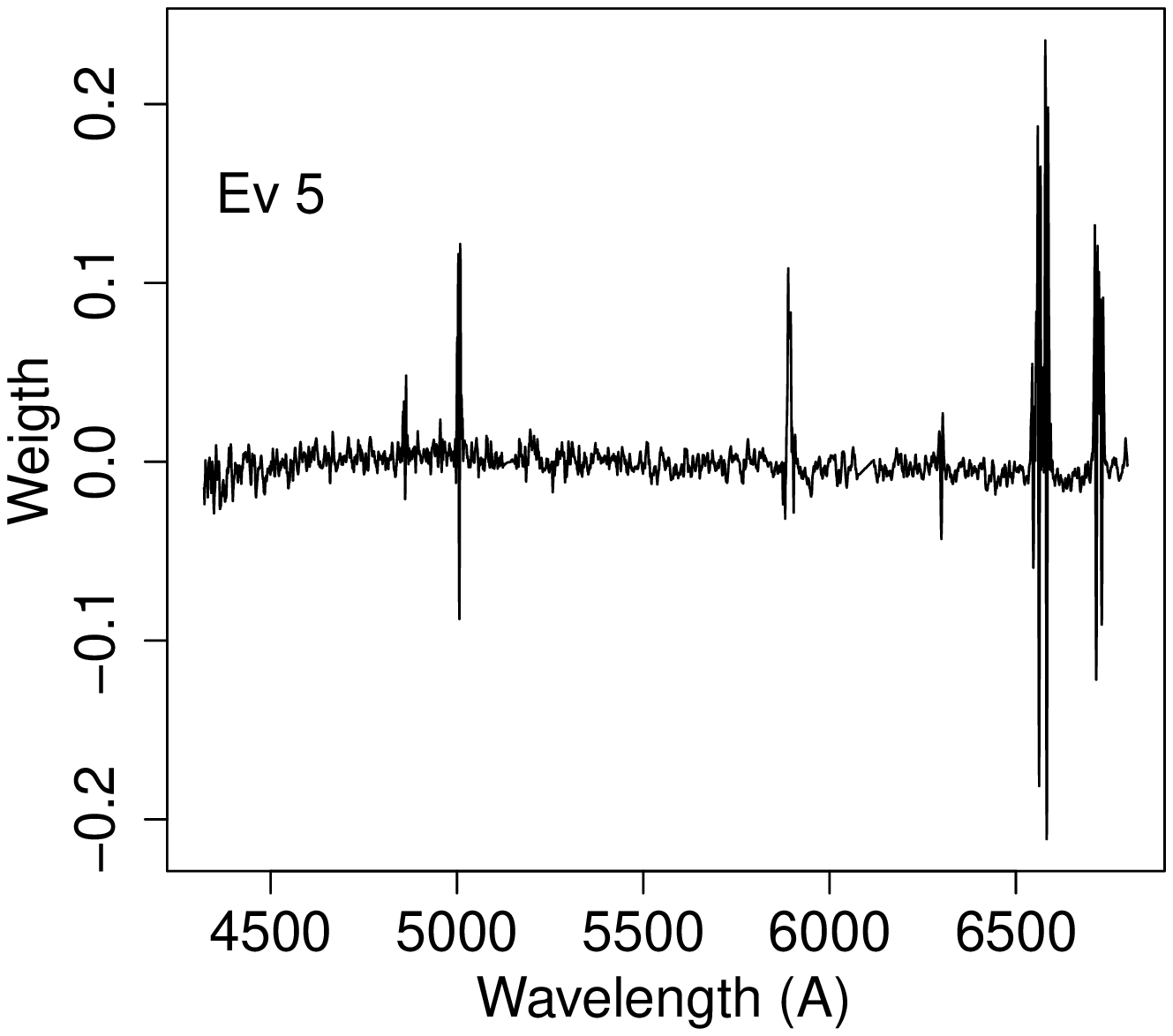}
\end{minipage}\hfill
\caption{Eigenvectors (Ev2 to Ev5) and tomograms of principal components 2-5 of the data cube of NGC 7097. In tomograms 3 and 5, ``black'' means correlation and ``white'' means anticorrelation. In tomograms 2 and 4, we display only the tip of the positive correlations.} 
\label{results_N7097}
\end{center}
\end{figure}

We will illustrate the method of PCA Tomography by applying it to our data of the galaxy NGC 7097 and interpreting the first 5 principal components in terms of physical properties. The first eigenvector, which explains 99.53\% of the data cube variance is shown in Figure \ref{Eigenvec1_fig} and is very similar to the integrated spectrum of the field of view (see Figure \ref{espec_total_av2eav4_N7097}). The added contribution of eigenvectors 2-5 explain 0.46\% of the variance in the data cube. This shows the great redundancy of this type of data and has relevance in discriminating non-redundant information. By removing the effects of the strongest correlations one can look to the less significant ones. The tomograms and eigenvectors and 2-5 are shown in Figure \ref{results_N7097}. 

Eigenvector 2, explaining 0.38\% of the variance, reveals correlations between features associated with transitions of H$\beta$, [O III] $\lambda \lambda$4959\AA , 5007\AA , [O I] $\lambda$6300\AA , H$\alpha$, [N II] $\lambda \lambda$6548\AA , 6583\AA\ and [S II] $\lambda \lambda$6714\AA , 6732\AA , with relative intensity typical of LINERs \citep{1980A&A....87..152H}. These characteristics are also correlated with a red continuum and interstellar absorption of Na I $\lambda$5891\AA. The tomogram shows the location of the object.

\begin{figure}[h!]
	\begin{center}
	\includegraphics[height=8cm, width=8cm, angle=0]{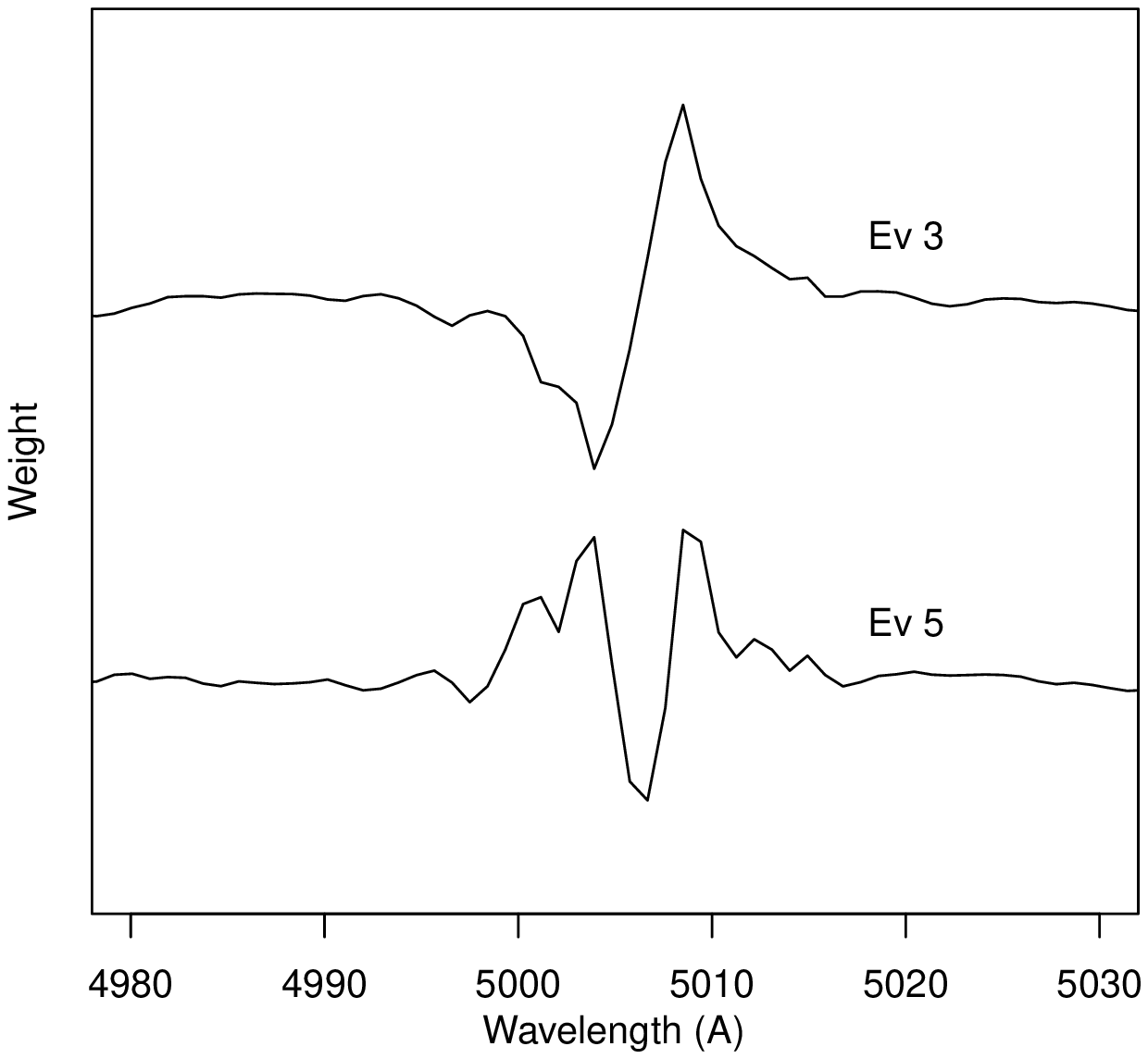}
	\caption{The expanded eigenvectors 3 (Ev 3 - top) and 5 (Ev 5 - bottom) for the wavelength region near the [O III]5007\AA\ line. These characteristics are interpreted as the anti-correlation produced by a rotating disk (in eigenvector 3) and that between the disk (wings of the lines) and the bi-cones (core of the lines) (in eigenvector 5).}
	\label{Av2_Av5_together}
	\end{center}
\end{figure}

In eigenvector 3, explaining 0.05\%, we observe an anti-correlation between the red and blue wings of the features associated to the emission lines (also see Figure \ref{Av2_Av5_together}). The tomogram reveals that the blue (approaching) material is in the south while
the red (receding) material is in the north. The position angle (PA) is $-10^o$. This is a signature of rotating gas, which is also seen in the galaxies NGC 4736 \citep{2009MNRAS.395...64S} and M 81 \citep{2011MNRAS.tmp...62S}.

In the fourth eigenvector, with 0.02\%, of the variance, there is again a correlation between emission line features typical of a LINER, but this time correlated with a blue continuum and the absence of interstellar absorption of Na I. The radial velocities in eigenvectors 2 and 4, estimated from the [O I] $\lambda$6300\AA\ transition, differ by less than 10 km/s, well below the spectral resolution, which is 90 km/s. 

The fifth eigenvector, with 0.01\% of the variance presents an eigenspectrum in which there is an anti-correlation between the low-velocity line features and their blue and red wings (see Figures \ref{results_N7097} and \ref{Av2_Av5_together}). The fifth tomogram reveals a low-velocity extended correlation that is located along the axis of rotation of the gas (perpendicular to the disk). The region associated with the line wings, for the high rotation velocity, corresponds to the disk, as seen in the third eigenvector. We interpret the low velocity region as the ionization bi-cone. 

How far can we look to the eigenvector before running into noise? We have made the ``scree test'' (see Figure 1 in \citealt{2009MNRAS.395...64S} for an explanation) and found that noise dominates all eigenvectors above number 7. Eigenvalue 8 is 0.0017\%, 10 times smaller than eigenvalue 5.

The LINER appears twice, in components 2 and 4. It is possible to reconstruct a data cube removing eigenvector 1 using the technique of  ``feature suppression'' (see section 6 of \citealt{2009MNRAS.395...64S}). Performing the suppression allows us to extract the spectrum of the central region, shown in Figure \ref{espec_total_av2eav4_N7097}. This illustrates, using only PCA, how one can isolate the spectrum of a low luminosity AGN decontaminated from the stellar and nebular emission. In Figure \ref{espec_total_av2eav4_N7097} we see that the Balmer lines have a weak broad component. This object is, therefore, of type 1 (sometimes also classified as type 1.8, given the weakness of the broad H$\alpha$, or $L_b$). The relative intensities of the narrow lines are consistent
with LINER \citep{1980A&A....87..152H}; we measured [O III] $\lambda$5009\AA\ / H$\beta$ $\sim$ 1.5, [N II] $\lambda$6584\AA\ / H$\alpha$ $\sim$ 1.2 (compare with \citealt{1997A&A...323..349P} and \citealt{1986AJ.....91.1062P}) and [O I] $\lambda$6300\AA\ / H$\alpha$ $\sim$ 0.34. For typical LINERs, [O III] $\lambda$5009\AA\ / H$\beta < 3$, [N II] $\lambda$6584\AA\ / H$\alpha > 0.5$ and [O I] $\lambda$6300\AA\ / H$\alpha > 0.05$ \citep{2006agna.book.....O}. From the original spectrum, we estimate the reddened H$\alpha$ luminosity as $5.0\pm1.0\times 10^{37} \rm erg\;s^{-1}$. We will not attempt to estimate the reddening as this will be done in more detail in a forthcoming paper.

\section{DISCUSSION AND CONCLUSIONS} \label{conclusions}

\begin{figure}[h!]
\includegraphics[height=5.30cm,width=7.80cm]{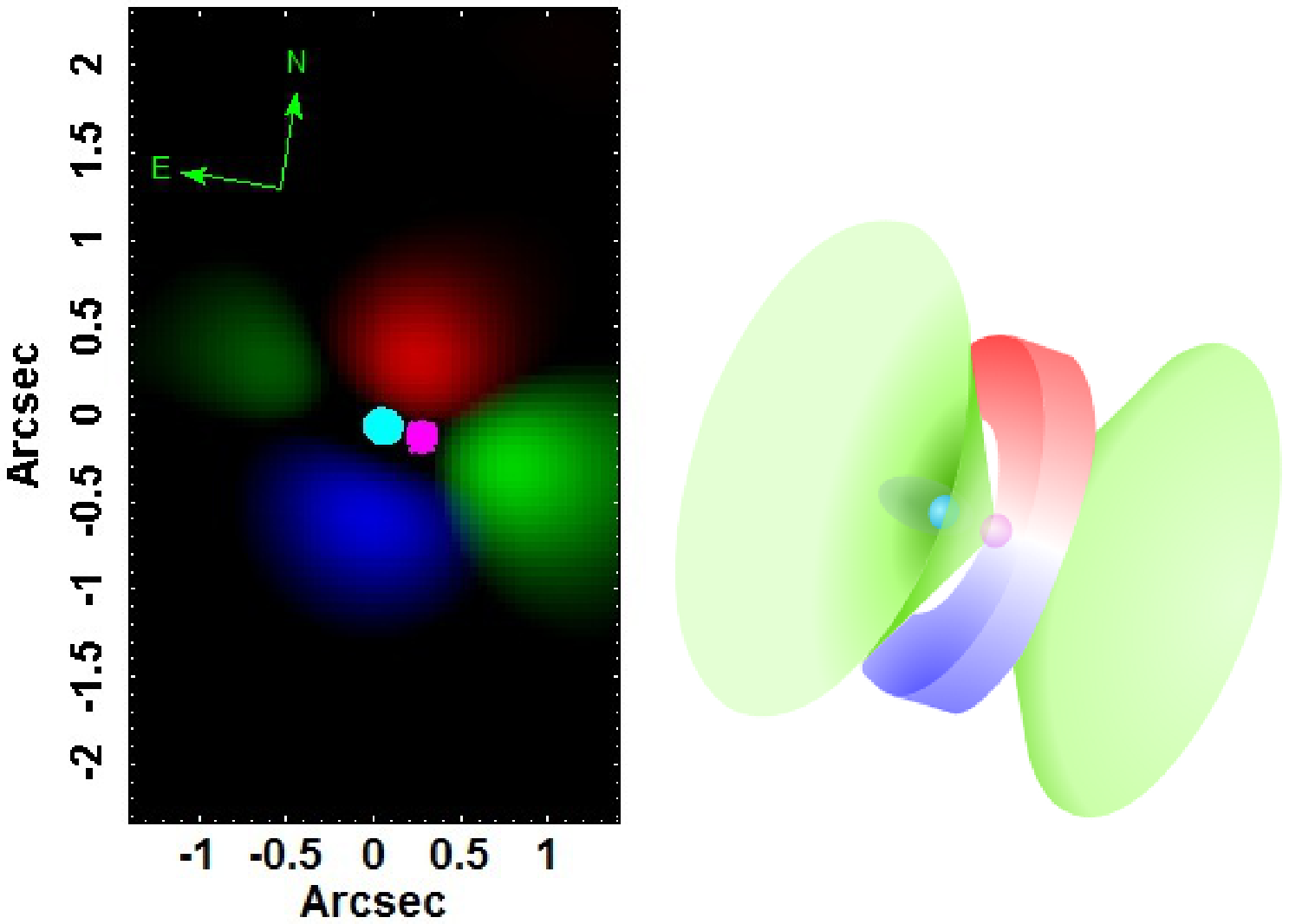}
\caption{Tomograms 2, 3, 4 and 5. The left image shows the combination of tomograms 2 (magenta), 3 (red=positive; blue=negative), 4 (cyan) and 5 (green). The extended emission in blue and red corresponds to the rotating disk of gas around the AGN. The small dots correspond to the emission of the AGN seen directly (magenta) and reflected in the ionization cone (cyan). The emission in green represents the bi-cone ionization. Right: Model of the central region of the galaxy NGC 7097. } 
\label{figura_modelo}
\end{figure}

We have seen that the eigenvectors and tomograms 2, 3, 4 and 5 can be interpreted as physical phenomena with distinct spectral properties. We summarize tomograms 2-5 in Figure \ref{figura_modelo}. On the left we see each one with a specific color (as indicated) and on the right we show a sketch that illustrates our interpretation. It seems clear that the LINER is seen twice: once directly through the disk of gas and dust (Ev 2), manifested by the heavily extincted continuum spectral slope and by interstellar neutral gas absorption. This same AGN is also seen (Ev 4) scattered by free electrons existing in the ionization cone; dust particles could also be present and produce additional scattering, especially in the blue. As we do not have reliable information on the absolute spectral behavior of the reflected continuum, we cannot say whether the scattering is independent of wavelength (as expected by electron scattering) or blue (if the scattering is produced by dust). This scattered image is less subjected to dust reddening or interstellar absorption by the disk, at least not as much as the object seen directly. 

The gaseous disk, identified in eigenvector 3, has a position angle of $PA= -10^o$; it is co-aligned with the disk described by \citet{1986ApJ...305..136C} and \citet{1993A&A...280..409B} which is $\sim$ 15 times larger.
 
We identify eigenvector 5 as representing the bi-cone in anti-correlation with the disk. The profile has a correlation in the low velocity core anticorrelated with the wings. Spatially, the wings are associated with the region corresponding to the disk (EV3). The central core in the eigenspectrum (see Figures \ref{results_N7097} and \ref{Av2_Av5_together}), seen in anti-correlation with the wings, is correlated with a structure in the respective tomogram that is perpendicular to the gaseous disk-looking, just like an ionization bi-cone. This disk-cone anti-correlation is expressed because the disk is quite highly inclined, presenting high velocities while the cones are less inclined with low intrinsic velocities. In agreement with this interpretation, an image of the [O III]$\lambda$5007\AA\ shows a weak extended emission along the near-side cone. One problem with this interpretation is that the far-side cone is slightly brighter than the near-side. We have to assume that this is an intrinsic asymmetry. Although this is not necessarily the only possible interpretation for the eigenvector 5, it seems to be self-consistent and is the only one we could find. The existence of this ionization bi-cone has not been reported previously for this galaxy.
In drawing the disk shown in Figure \ref{figura_modelo}, we have assumed the inclination angle determined by \citet{1997A&A...323..349P}, $67^o$, and postulated a cone half-opening angle of $57^o$. An opening angle much different from this would not produce two such close AGN images as observed.

The separation between the two images of the AGN is about 0.2'' while the seeing was $\sim$ 1'' (Figure \ref{figura_modelo}). This is possible because the PCA is a filtering technique.  Each tomogram produces an image of correlations and, therefore, can be interpreted representing the data in distinct ``filters''. This does not produce super-resolution but can image phenomena with angular  separations smaller than the resolution. The basic reason for this is that correlations are present below the seeing limit and not destroyed by the random variations in the PSF. We predict that the blue dot in Figure \ref{figura_modelo}  should have polarized light because of its scattered nature.

In conclusion, we presented a new method of locating and extracting scattered light from AGN; this does not use the well known method of spectropolarimetry but uses PCA analysis combined with 3D spectroscopy.

\acknowledgments {We would like to thank Alex Carciofi and J. E. Bjorkman for critically reading the manuscript. We also thank Felipe Andrade Oliveira for helping in the ilustration. T.V.R. and R.B.M. thank FAPESP for support under grants 2008/06988-0 (T.V.R) and 2008/11087-1 (R.B.M.). We would like to thank an anonymous referee for his valuable suggestions that improved the quality of this Letter.}

\end{document}